\documentclass{aastex631}

\shorttitle{12 Contact Binaries}
\shortauthors{Wadhwa et al.}
\graphicspath{{./}{figures/}}

\begin{document}

\title{A Study of Twelve Potential Merger Candidate Contact Binary Systems}

\author[0000-0002-7011-7541]{Surjit S. Wadhwa}
\affiliation{School of Science, Western Sydney University,\\ Locked Bag 1797, Penrith, NSW 2751, Australia.\\}
\author[0000-0002-8036-4132]{Bojan Arbutina}
\affiliation{Department of Astronomy, Faculty of Mathematics, University of Belgrade, Studentski trg 16, 11000 Belgrade, Serbia.\\}
\author[0000-0002-9931-5162]{Nick F. H.  Tothill}
\affiliation{School of Science, Western Sydney University,\\ Locked Bag 1797, Penrith, NSW 2751, Australia.\\}
\author[0000-0002-4990-9288]{Miroslav D. Filipovi\'c}
\affiliation{School of Science, Western Sydney University,\\ Locked Bag 1797, Penrith, NSW 2751, Australia.\\}
\author[0000-0001-9677-1499]{Ain Y. De Horta}
\affiliation{School of Science, Western Sydney University,\\ Locked Bag 1797, Penrith, NSW 2751, Australia.\\}

\author[0000-0001-8535-7807]{Jelena Petrovi\'c}
\affiliation{Astronomical Observatory, Volgina 7, 11060 Belgrade, Serbia\\}
\author[ 0000-0001-9392-6678]{Gojko Djura\v sevi\'c}
\affiliation{Astronomical Observatory, Volgina 7, 11060 Belgrade, Serbia\\}



\begin{abstract}

Photometric observations and analysis of twelve previously poorly studied contact binary systems is presented. All show total eclipses and have extremely low mass ratios ranging from 0.072 to 0.15.  Also, all show characteristics of orbital instability with mass ratios within the theoretical orbital instability range. Although none demonstrate a significant O'Connell effect at least nine of the systems have other indicators of increased chromospheric and magnetic activity. 
\end{abstract}

\keywords{Contact Binary, Low Mass Ratio, Red Nova}


\section{Introduction} \label{sec:intro}

As the number of identified contact binary systems grows with new discoveries from an ever growing number of sky surveys, the interest in their investigation has also increased. The confirmation of Nova 2008 Sco (V1390 Sco) as a red nova resulting from the merger of contact binary components \citep{2011A&A...528A.114T} has also intensified interest in the investigation of orbital stability of contact binary systems \citep{2021MNRAS.501..229W, 2021MNRAS.502.2879G, 2022MNRAS.512.1244C, 2023MNRAS.519.5760L}. Theoretical models of contact binary evolution predict a likely merger event leading to a rapidly rotating cool giant star as the final outcome \citep{2003ApJ...582L.105S, 2012JASS...29..145E, 2011A&A...531A..18S, 1977MNRAS.179..359R}. Although the predicted rate of Galactic merger events is high at once every 2 to 3 years, brighter events suitable for study are likely to be much rarer at once every 10 or more years \citep{2014MNRAS.443.1319K}. 

Investigators such \citet{1995ApJ...438..887R}, \citet{2007ApJ...662..596L} and \citet{2007MNRAS.377.1635A, 2009MNRAS.394..501A} have clearly shown that merger events are likely to occur when the mass ratio of the components is low. More recently \citet{2021MNRAS.501..229W} linked the instability parameters to the mass of the primary ($M_1)$ and concluded that there is not one global minimum mass ratio at which merger will take place, rather the instability mass ratio can range from below 0.05 to above 0.2 for systems where $0.6M_{\sun}<M_1<1.4M_{\sun}$. Although a number of studies, each with a large (10 or greater) number of contact binaries with extremely low mass ratios, have recently been reported they did not identify any system that meet the instability criteria \citep{2021MNRAS.502.2879G, 2022MNRAS.512.1244C, 2022AJ....164..202L, 2023MNRAS.519.5760L}. \citet{2022JApA...43...94W} recently linked the amplitude of totally eclipsing contact binaries with the mass ratio which when used in conjunction with estimates of the mass of the primary can result in the rapid identification of potentially unstable systems. They identified over 40 new potential merger candidates based on the analysis of survey photometric data. In this study we perform follow up ground based observations on nine systems identified by \citep{2022JApA...43...94W} and three other systems identified from the All Sky Automated Survey - Super Nova (ASAS-SN) \citep{2014ApJ...788...48S, 2020MNRAS.491...13J} using the criteria described in \citet{2022JApA...43...94W}. Basic identification data for each system is presented in Table 1. Almost all ground based surveys have been carried out using either small aperture telescopes or telephoto lenses. The angular resolution is therefore quite poor ranging from 9" to over 60" \citep{2017MNRAS.469.3688D, 2014ApJ...788...48S, 2020MNRAS.491...13J, 2021MNRAS.502.1299T, 2002AcA....52..397P}. The risk of potential blending of the survey light curves was highlighted by \citet{2022arXiv221208209W} where a system thought to be of extreme low mass ratio and a potential merger candidate is in fact a high mass ratio system with no indications of orbital instability. Each system reported here was checked against $Gaia$ EDR 3 \citep{2022A&A...658A..91A, 2022arXiv220800211G} to ensure there was no other star system brighter than 17th magnitude within 10" of the target. We perform light curve analysis and compare our results with reported data from survey analysis, in addition, we show that all systems described demonstrate features of orbital instability, have secondaries that are denser and most show evidence of chromospheric activity in the absence of star spots.

\begin{table}

   \centering

   \begin{tabular}{|l|l|l|l|}
    \hline
        \hfil Name &\hfil Abbreviation &\hfil Comparison Star &\hfil Check Star \\ 
        \hline
        \hfil ASAS J045814+0643.1 & \hfil A0458 &\hfil TYC 97-681-1 & \hfil TYC 97-273-1 \\
        
         \hfil ASAS J051459-7356.3 &\hfil A0514   &\hfil TYC 9174-356-1  &\hfil 2MASS 05140657-7351334\\ 
          \hfil ASAS J100101-7958.6 &\hfil A1001   &\hfil TYC 9404-105-1  &\hfil 2MASS	09593687-7954508 \\ 
          \hfil V396 Lup &\hfil V396 Lup  &\hfil TYC 7851-983-1 &\hfil 2MASS16031759-3751375\\ 
          \hfil ASAS J170715-5118.7 &\hfil A1707 &\hfil TYC 8340-637-1 &\hfil2MASS 17072965-5118000 \\ 
           \hfil ASAS J184644-2736.4 &\hfil A1846  & \hfil TYC 6867-2234-1 &\hfil TYC 6867-2250-1 \\
           \hfil ASAS J202231-4452.5 & \hfil A2022 &\hfil TYC 7961-1064-1 & \hfil TYC 7961-842-1 \\
        \hfil ASAS J204452+0622.6 &\hfil A2044  &\hfil 2MASS 20445329+0619549  &\hfil 2MASS 20444317+0619507 \\ 
        \hfil ASAS J213219-5351.6 &\hfil A2132  &\hfil UCAC2 8680692  &\hfil UCAC4 181-218564 \\
        \hfil SSS-J221327.1-445401 &\hfil S2213   &\hfil UCAC2 12815832  &\hfil 2MASS 22140108-4453567\\
          \hfil ASAS J225826-2603.6 &\hfil A2258  &\hfil TYC 6974-1080-1 &\hfil TYC 6974-880-1\\ 
          \hfil ASAS J234823-4054.7 &\hfil A2348 &\hfil TYC 8018-55-1 &\hfil 2MASS	23480833-4046082 \\ \hline
           
    \end{tabular}
    \caption{Name, Abbreviations (used in all other tables and text) along with comparison and check stars for 12 contact binary systems}
    \end{table}

\section {Photometric Observations, Absolute magnitudes and Mass of the Primary Component}
\subsection{Photometric Observations}
We acquired dual band (V and R) images of 12 contact binary systems from April 2020 to January 2023 using either the 0.6m Western Sydney University (WSU) telescope equipped with a cooled SBIG 8300 CCD camera and standard Johnson $BVR$ filters or the 0.4m telescopes of the Las Cumbres Observatory (LCO) network equipped with SBIG STL-6303 CCD camera and Bessel V,B and Sloan r' filters. In addition to V and R band images, B band images were obtained only during eclipses to document $B-V$ magnitude for each system. The LCO network automatically calibrates all images while the images obtained at the WSU were calibrated with multiple flat, dark and bias frames. Differential photometry was performed for each system using the AstroImageJ \citep{2017AJ....153...77C} software package. All observations with software reported error of greater than 0.01 magnitude were excluded. Observation dates, number of observations, exposure times, maximum V band brightness, V Band amplitude and $B-V$ magnitude are summarised in Table 2.

\begin{table}

   \centering

   \begin{tabular}{|l|l|l|l|l|l|l|l|l|}
    \hline
        \hfil Name &\hfil Obs Date &\hfil Obs (V,R) &\hfil Exp Times (V,R)&\hfil Max (V) &\hfil Ampl (V) &\hfil B-V&\hfil $M_{V1}$&\hfil $M_1/M_{\sun}$ \\
        \hline
        \hfil A0458 &\hfil 11/21 - 01/23 &\hfil 285, 350  &\hfil 45s,45s&\hfil 11.76&\hfil 0.26&\hfil 0.46&\hfil $3.88\pm0.01$&$1.17\pm0.02$ \\
        
        \hfil A0514   &\hfil 01/22 - 10/22 &\hfil 320, 300 &\hfil 45s,45s&\hfil 11.71&\hfil 0.37&\hfil 0.71&\hfil $4.90\pm0.02$&$0.98\pm0.01$ \\ 
         \hfil A1001   &\hfil 04/20 - 04/20  &\hfil 373, 375&\hfil 40s,40s&\hfil 11.47&\hfil 0.33&\hfil 0.91&\hfil $5.78\pm0.01$ &$0.83\pm0.01$ \\ 
          \hfil V396 Lup &\hfil06/21 - 07/21  &\hfil 670, 240 &\hfil 30s,25s&\hfil 10.88&\hfil 0.36&\hfil 0.83&\hfil $5.20\pm0.02$&$0.95\pm0.01$ \\ 
          \hfil A1707 &\hfil 07/21 - 08/21 &\hfil510, 500&\hfil 40s,35s&\hfil 11.19&\hfil 0.44&\hfil 0.55&\hfil $3.67\pm0.02$&$1.22\pm0.01$ \\ 
           \hfil A1846  & \hfil 08/21 - 08/21 &\hfil300, 410&\hfil 40s,40s&\hfil 11.78&\hfil 0.39&\hfil 0.78&\hfil $5.43\pm0.02$&$0.90\pm0.01$ \\
           \hfil A2022 &\hfil 08/21 - 08/21 & \hfil 550, 300&\hfil 45s,45s&\hfil 12.09&\hfil 0.31&\hfil 0.53&\hfil $4.38\pm0.01$&$1.11\pm0.01$ \\
        \hfil A2044  &\hfil 08/22 - 10/22  &\hfil 440, 280&\hfil 50s,45s&\hfil 12.67&\hfil 0.30&\hfil 0.62&\hfil $4.18\pm0.02$&$1.13\pm0.01$  \\ 
        \hfil A2132  &\hfil 10/21 - 09/22 &\hfil 255, 250&\hfil 55s,55s&\hfil 12.89&\hfil 0.28&\hfil 0.55&\hfil $4.48\pm0.02$&$1.04\pm0.01$  \\
        \hfil S2213   &\hfil 09/21 - 09/21  &\hfil 250, 300&\hfil 45s,45s&\hfil 12.49&\hfil 0.27&\hfil 0.55&\hfil $4.05\pm0.02$&$1.15\pm0.02$ \\ 
          \hfil A2258  &\hfil 09/21 - 10/21 &\hfil 255, 330&\hfil 35s,35s&\hfil 11.49&\hfil 0.26&\hfil 0.60&\hfil $4.21\pm0.02$&$1.12\pm0.01$\\ 
          \hfil A2348 &\hfil 11/21 - 08/22 &\hfil 350, 350&\hfil 50s,50s&\hfil 12.83&\hfil 0.30&\hfil 0.54&\hfil $4.16\pm0.03$ &$1.11\pm0.02$\\ \hline
           
    \end{tabular}
    \caption{Observation dates, number of observations, exposure times, light curve parameters and spectral classification of the twelve reported contact binary systems. A1001, V396 Lup, A1707 and A1846 were observed with the WSU telescope others through the LCO telescope network. $^*$ = Spectral classification from published literature.}
    \end{table}

Based on the times of minima observed and available V band photometric data from the ASAS-SN and All Sky Automated Survey (ASAS) \citep{2002AcA....52..397P} we updated the orbital elements as summerised in Table 3.

\begin{table}

   \centering

   \begin{tabular}{|l|l|l|}
    \hline
        \hfil Name &\hfil Epoch (HJD) &\hfil Period (d) \\
        \hline
        \hfil A0458 &\hfil $2459543.900671\pm0.000728$ &\hfil$ 0.333482\pm0.000012$ \\
        
        \hfil A0514   &\hfil $2459594.108439\pm0.000520$ &\hfil $0.345729\pm0.000085$  \\ 
         \hfil A1001   &\hfil $2458961.971160\pm0.000920$  &\hfil $0.279133\pm0.000085$ \\ 
          \hfil V396 Lup &\hfil$2459372.012959\pm0.000420$  &\hfil $0.363248\pm0.000060$ \\ 
          \hfil A1707 &\hfil $2459430.984062\pm0.000511$ &\hfil$0.525901\pm0.000040$ \\ 
           \hfil A1846  & \hfil $2459439.005458\pm0.000225$ &\hfil$0.302853\pm0.000030$ \\
           \hfil A2022 &\hfil $2459429.668598\pm0.000986$ & \hfil $0.345011\pm0.000006$\\
        \hfil A2044  &\hfil $2459818.566466\pm0.000564$  &\hfil $0.370536\pm0.000029$ \\ 
        \hfil A2132  &\hfil $2459509.962001\pm0.000488$ &\hfil $0.316348\pm0.000058$ \\
        \hfil S2213   &\hfil $2459459.704234\pm0.000621$  &\hfil $0.369894\pm0.000005$\\ 
          \hfil A2258  &\hfil $2459436.659418\pm0.000267$ &\hfil $0.327648\pm0.000060$\\ 
          \hfil A2348 &\hfil $2459545.581557\pm0.000223$ &\hfil $0.347195\pm0.000052$\\ \hline
           
    \end{tabular}
    \caption{Updated orbital elements}
    \end{table}

\subsection{Absolute Magnitude of the Primary Component}
As all systems described have a very low mass ratio (see below) and all demonstrate total eclipses we are presented with a prospect of being able to estimate the absolute magnitude of the primary component using direct observations. The apparent magnitude of the secondary eclipse represents the apparent magnitude of the primary. If the distance is known, the absolute magnitude of the primary ($M_{V1}$) can be estimated as follows:

\begin{equation}
    M = m-5log_{10}d + 5
\end{equation}

Where M is the absolute magnitude, m is the apparent magnitude and d is the distance in parsecs.

The $GAIA$ mission \citep{2022A&A...658A..91A} provides highly accurate estimates of the distance particularly of nearby stars and as such makes distance based estimate of the absolute magnitude a reality. Interstellar extinction however must be taken into consideration when determining the apparent magnitude. As most dust maps report line of sight extinction to infinity and all our systems have distance estimations well below 1~kpc we corrected the extinction for distance as follows: Using \citet{2011ApJ...737..103S} dust maps we determined the line of sight reddening at infinity $E(B-V)_{\infty}$. We used the $Gaia$ distance to scale this value $E(B-V)_d$ using the equation \citep{2008MNRAS.384.1178B}:

\begin{equation}
    E(B-V)_d = E(B-V)_{\infty}\Bigg[1-\mathrm{exp}\bigg(-\frac{|d{\cdot \mathrm{sin}(b)}|}{h}\bigg)\bigg]
\end{equation}
In the equation $b$ is the galactic latitude of the system, $d$ is the $Gaia$ distance (in parsecs) and $h$ is the galactic scale height, taken as $h=125pc$ as per \citet{2008MNRAS.384.1178B}. The total extinction  ($A_V$) was then estimated as $A_V = E(B-V)_d\times3.1$.

The estimated absolute magnitude for each system is summarised in Table 2. The largest contributor to the estimated error in the absolute magnitude was the reported error in the distance estimate.

\subsection{Mass of the Primary Component}

One of the main aims of this report is to assess potential orbital instability of the reported contact binary systems. Recently,  \citet{2021MNRAS.501..229W} developed new relationships linking the mass of the primary ($M_1$), the degree of contact and the instability mass ratio ($q_{inst}$). One can estimate the range of mass ratios over which a system may become unstable if the mass of the primary is known. Direct estimation of the mass of the primary is not possible, one is therefore reliant on secondary observations and calibrations. Primary components of contact binary systems follow a main sequence profile \citep{2013MNRAS.430.2029Y}. For this study we use two methods to estimate the mass of the primary. The Two Micron All Sky Survey (2MASS) \citep{2006AJ....131.1163S} acquired simultaneous photometry in multiple infrared bands. As our first estimate of the mass of the primary we used the  2MASS $J-H$ magnitudes and the April 2022 update tables of \citet{2013ApJS..208....9P} for low mass ($0.6M_{\sun} < M_1 < 1.4M_{\sun}$) main sequence stars to interpolate the mass of the primary. Our distance based estimate is linked to the absolute magnitude of the primary determined above. Again we use the April 2022 update tables of \citet{2013ApJS..208....9P} for low mass ($0.6M_{\sun} < M_1 < 1.4M_{\sun}$) main sequence stars to interpolate the mass of the primary. We use the mean of the colour and distance based estimates as our adopted value for the mass of the primary. The recorded errors were propagated when calculating the instability mass ratio range (see below). Summary results for the mass of the primary are recorded in Table 2.

\section{Light Curve Analysis and the Mass Ratio}

Having established an estimate for the mass of the primary the next (and only) other parameter required to determine potential orbital stability is the mass ratio of the system. In the absence of radial velocity measurements light curve analysis of contact binary systems can be successfully carried out if total eclipses are present \citep{2005Ap&SS.296..221T}. One of the input parameters for such analysis is the temperature of the primary component ($T_1$). There is no standard method for assigning the value for the temperature of the primary. Colour based estimations have been the mainstream, however these have proven troublesome, for example the ViziR database reports a range of over 3000K for one of our targets (V396 Lup). Given the lack of a standard approach many investigators  \citep[see e.g.][]{2018PASJ...70...87Z, 2022PASJ...74.1421C, 2022MNRAS.517.1928G} are now moving to low resolution spectra (where available) to aid in determining the primary's temperature. To this end for this study we acquired spectra using the LCO network of 2m telescopes equipped with FLOYDS spectrograph, a cross-dispersed, spectrograph with variable resolution of R=400 to R=700 for 10 of the systems. The reduction pipeline for the FLOYDS spectral data is fully automated and described in detail on the LCO website (https://lco.global/documentation/data/floyds-pipeline/ - accessed 14/1/2023). For A0514 we used the published spectral data \citep{1993yCat.3135....0C}. We were unable to obtain a spectrum for A2044 due to technical reasons during the allotted time. All FLOYDS spectra were visually compared to standard library spectra of main sequence stars \citep{1984ApJS...56..257J, 1998PASP..110..863P} to determine the spectral class for each system. Representative FLOYDS spectra and matching library spectra are shown in Figure 1.

For this study we use the mean ($\pm{SD}$) of four estimates for the temperature of the primary. We estimate effective temperature of the primary based on our extinction corrected value of $B-V$, the observed spectral class, the 2MASS $J-H$ colour and the reported value from the GAIA third data release all adjusted to the nearest 10. For consistency we use the April 2022 update tables of \citet{2013ApJS..208....9P} for colour and spectral calibrations. Summary of the spectral class, temperatures and the adopted mean value of the four calibrations is presented in Table 4. In the case of A2044 where no spectral observations were available we adopt the mean of the two colour calibrations and the GAIA estimate.

\begin{table}

\centering
  \centering

   \begin{tabular}{|l|l|l|l|l|l|l|}
   \hline
        \hfil Name &\hfil $B-V (K)$ &\hfil $J-H (K)$ &\hfil $Sp (K)$ &\hfil $GAIA (K)$&\hfil $Mean\pm SD$&\hfil Sp Class \\
        \hline
        \hfil A0458& \hfil 6650 &\hfil 6060 & \hfil 7020&\hfil7110&\hfil$6710\pm410$&\hfil F1 \\
        
        \hfil A0514 &\hfil 5810   &\hfil 5600  &\hfil 5930&\hfil 5310&\hfil $5660\pm240$&\hfil G0$^*$\\ 
         \hfil A1001&\hfil 5140   &\hfil 4870  &\hfil 5380&\hfil 4990&\hfil $5100\pm190$&\hfil G9 \\ 
          \hfil V396 Lup &\hfil 5520  &\hfil 5550 &\hfil 6050&\hfil 5930&\hfil $5760\pm230$&\hfil F9\\ 
         \hfil A1707&\hfil 6610 &\hfil 6190 &\hfil6750 &\hfil 6550&\hfil $6530\pm210$&\hfil F3\\ 
         \hfil A1846&\hfil 5530  & \hfil 5370&\hfil 5550&\hfil 5080&\hfil $5380\pm190$&\hfil G7 \\
           \hfil A2022& \hfil 6280 &\hfil 6060 & \hfil5930&\hfil 6240&\hfil $6130\pm140$&\hfil G0  \\
        \hfil A2044&\hfil 6070  &\hfil 6060  &\hfil -&\hfil 5900&\hfil $6010\pm80$&\hfil - \\ 
       \hfil A2132&\hfil 6190  &\hfil 5710  &\hfil 6180&\hfil 6020&\hfil $6030\pm190$&\hfil F8 \\
        \hfil S2213&\hfil 6190   &\hfil 6060  &\hfil 6280&\hfil 5970&\hfil $6130\pm120$&\hfil F7\\
          \hfil A2258&\hfil 6000  &\hfil 5990 &\hfil 6050&\hfil 5910&\hfil $5990\pm50$&\hfil F9\\ 
          \hfil A2348&\hfil 6180 &\hfil 5930 &\hfil 6550&\hfil 6110&\hfil $6190\pm230$&\hfil F5 \\ \hline
           
    \end{tabular}
    \caption{Effective temperatures based on $B-V$, $J-H$, spectral class and GAIA. Mean value $\pm (SD)$ was the adopted value for light curve analysis. In the last column we record the spectral class for each system. $^*$ In the case of A0514 we use the published data \citep{1993yCat.3135....0C}.}
    \end{table}

The 2013 version of the Wilson-Devinney (WD) code with Kurucz atmospheres \citep{1990ApJ...356..613W, 1998ApJ...508..308K, 2021NewA...8601565N} was used to obtain the photometric solution for each system. As all systems show total eclipses accurate photometric solutions are possible and as there was no significant asymmetry at maximum brightness (i.e no significant O'Connell Effect) only unspotted solutions were obtained. As the effective temperature of the primary is less than 7200K in all cases, the gravity coefficients were set as $g_1 = g_2 = 0.32$, and bolometric albedoes as $A_1 = A_2 = 0.5$. Logarithmic \citep{2015IBVS.6134....1N} limb darkening coefficients were interpolated from the 2019 update of \citet{1993AJ....106.2096V}.

Simultaneous solutions were obtained for both the V and R band in all cases. Mass ratio ($q$) search grid method was used to find the best solution for a range of fixed mass ratios from 0.05 to 20. The initial search was performed with mass ratio increments of 0.1 up to $q=1$ and then 0.2 up to $q=10$ and 0.5 increments up to $q=20$. The search was then refined near the best solution in increments of 0.05 and finally in increments of 0.01. During the search procedure the orbital inclination ($i$), the surface potential, temperature of the secondary component ($T_2$) and the dimensionless luminosity of the primary ($L_1$) were the adjustable parameters and all of which were adjusted between each iteration. The iterations were carried out until the suggested adjustment for all adjustable parameters was less than the reported standard deviation. During the last iteration the mass ratio was also made a adjustable parameter. 

To determine the effects of the variable temperature of the primary each system was further modelled with mass ratio increments of 0.01 around the mean temperature solution with the temperature of the primary adjusted both to the upper and lower limits reported in Table 4. As has been well established the light curve shape of contact binaries is almost completely due to the system geometry with the degree of contact, mass ratio and inclination being the major contributors \citep{1993PASP..105.1433R, 2001AJ....122.1007R}. The shape of contact binary light curves places a tight constrain on the component temperature ratio ($T_2/T_1$) but not on individual component temperatures \citep{1993PASP..105.1433R, 2020ApJS..247...50S}. With change in $T_1$ we expected change in $T_2$ but no significant change in the geometric parameters. This proved to be the case with the light curve solution converging to the same mass ratio regardless of the input value of $T_1$ except in the case of A1846 where the mean $T_1$ solution converged at $q=0.151$ and the higher $T_1$ solution converged at $q=0.150$. Similarly the inclination variation between high, mean and low values for $T_1$ were within 0.5$^\circ$ and within the reported standard deviations reported by the WD software. The dimensionless potential and the geometric mean of the fractional radii again varied minimally. Although the errors in geometric parameters were small with change in temperature of the primary they were propagated to the mean solution reported. Apart from the mass ratio and fractional radii (reported to two decimal places) no other parameter resulting from modelling of the light curve is further used in this study. We report light curve solution parameters pertinent to this study in Table 5 with errors propagated relative to the input errors in $T_1$. Fitted and observed light curves are illustrated in Figure 2.

\begin{figure}[!ht]
    \label{fig:F1}
	\includegraphics[width=\textwidth]{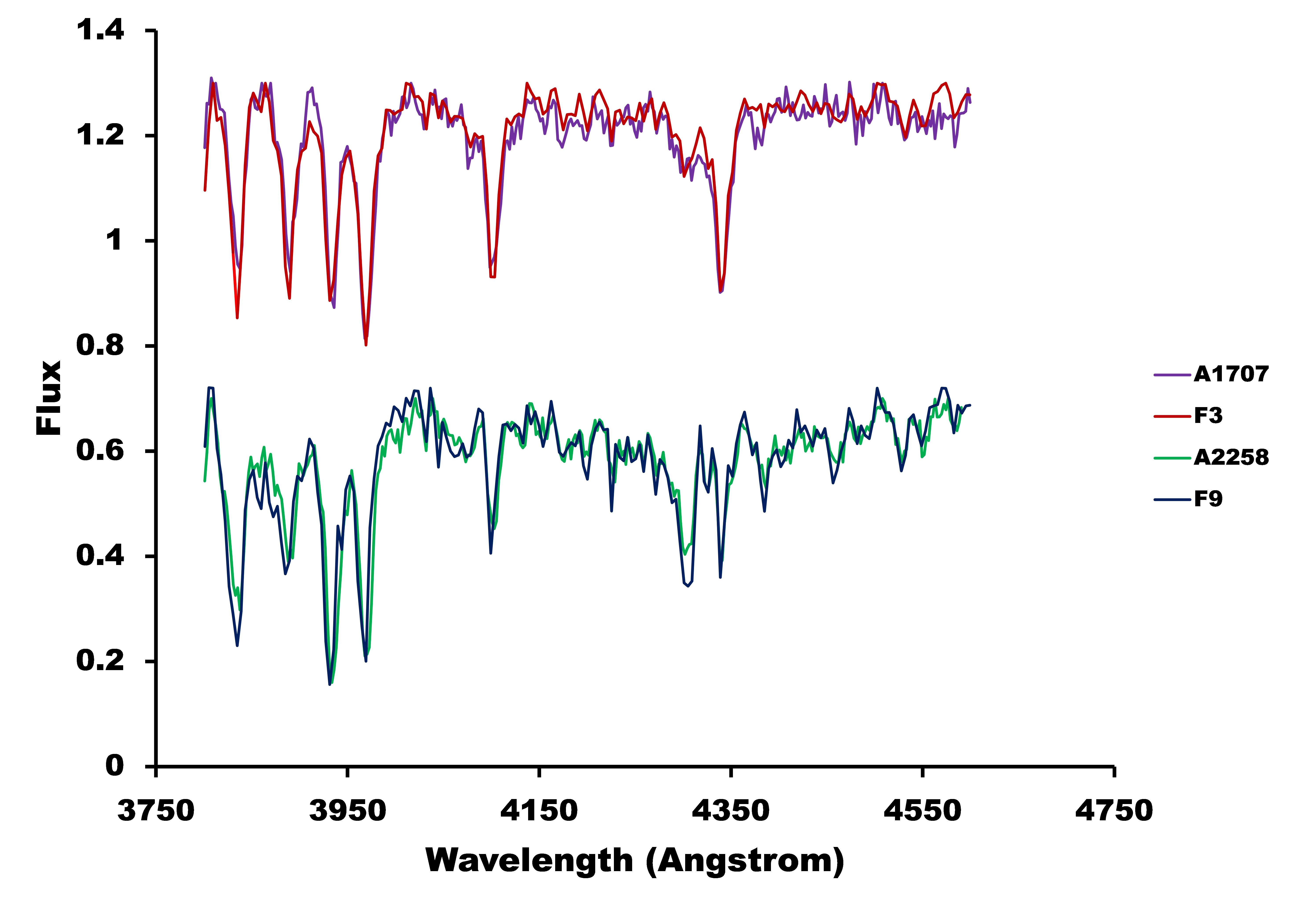}
    \caption{The low resolution FLOYDS spectra were visually matched to standard main sequence star spectra to determine the spectral class for 10 of the systems.}
    \end{figure}

\section{Orbital stability}
Orbital stability and potential merger of contact binary systems has received considerable attention recently \citep{2021MNRAS.501..229W, 2022MNRAS.512.1244C, 2023MNRAS.519.5760L}. \citet{2021MNRAS.501..229W} developed new relationships linking the mass of the primary, the degree of contact and the instability mass ratio ($q_{inst}$). Following their methodology, and adopting the values of the gyration radii of the components as described, we calculated $q_{inst}$ for each system for fill-out  factor values of 0 and 1 (Table 5). This provides a range of mass ratios where orbital instability is likely. As noted by \citet{2022MNRAS.512.1244C} a small uncertainty in the estimation of the mass of the primary propagates to a significant uncertainty in the estimation of orbital instability parameters. This is confirmed by this study where the mean mass of the primary components of the systems presented in this study is $1.06M_{\sun}$. At this mass the instability mass ratio range for fill-out factor 0 to 1 extends from 0.094 to 0.109. A 10\% change in the mass of the primary (either above or below) results in a greater than 17\% change in the instability mass ratio range. Thus for the purpose of this study we consider any system where the mass ratio is within 17\% of the maximum instability mass ratio to be potentially unstable.  Based on this criteria ten of the twelve systems lie within the instability mass ratio range taking into account reported errors. One (A1846) is within 7\% and one (A2348) 17\% of the maximum instability mass ratio. We note that the distance estimation of A2348 has a reported error near 10\% so we consider it reasonable to include the system as potentially unstable. The instability mass ratio ranges are summarised in Table 5.

Manual analysis of the survey photometric data for 9 of the systems presented here was reported by \citet{2022JApA...43...94W}. The survey data mass ratio estimate was within 10\% of the mass ratio based on dedicated observations except in the case of A2044 and A2132 where the survey data estimates of the mass ratio were 20\% and 12\%,  respectively, below the values reported here.The findings confirm that generally survey photometric data provides a good starting point in the selection of potential low mass ratio systems and merger candidates. \citet{2022JApA...43...94W} found that all 9 of the common systems could be regarded as potential merger candidates as their estimation of the mass of the primary was based on $J-H$ calibration. Looking at the two candidates in our report A1846 and A2348 which are just outside the instability range, have $J-H$ calibrated mass of the primary, 0.89 and 1.07, respectively, both lower than the combined with distance estimation used in this study. If the single colour estimation was used than the 2 would fall within the instability range.

\begin{figure}[!ht]
    \label{fig:F2}
	\includegraphics[width=\textwidth]{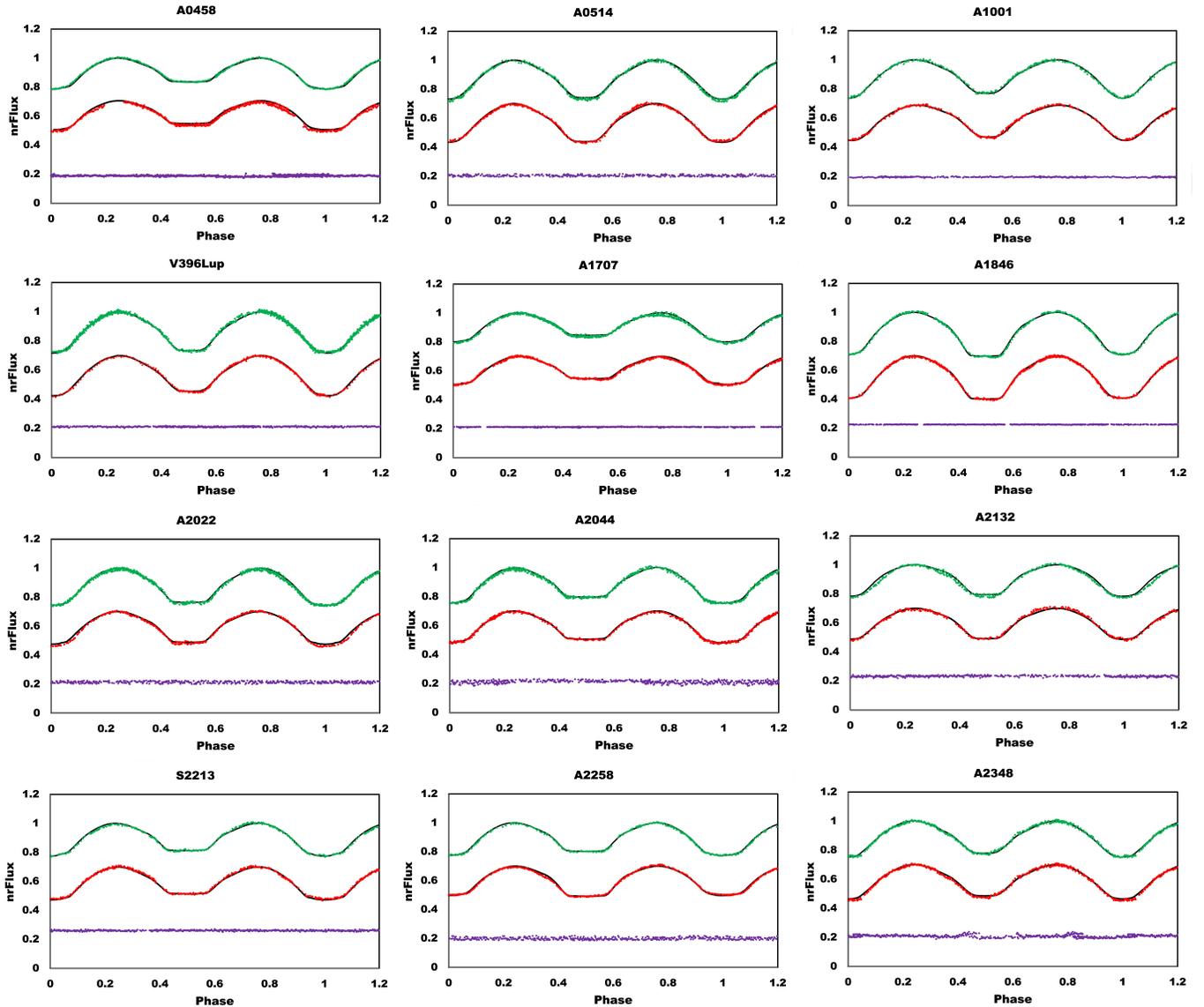}
    \caption{Observed and modelled light curves for the 12 systems. The green and red light curves represent the V and R(r') bands while the purple curve represents the check star. The black line represents the WD modelled curve. The vertical axis labelled nrFlux represents the normalised flux which has been arbitrarily shifted vertically for clarity}
    \end{figure}

\begin{table*}

   \centering

   \begin{tabular}{|l|l|l|l|l|l|l|}
    \hline
        \hfil Name &\hfil$T_1 (K)$ &\hfil $T_2 (K)$ &\hfil Incl $(^\circ)$&\hfil Mass Ratio ($q$) &\hfil Fr Radii ($r_{1,2}$) &\hfil $q_{inst}$ Range \\
        \hline
        \hfil A0458 &\hfil $6710\pm410$ &\hfil $6190\pm380$  &\hfil $88.4\pm0.8$&\hfil $0.089\pm0.001$&\hfil $0.60\pm0.01$, $0.21\pm0.01$&\hfil $0.079\pm0.003$ - $0.09\pm0.003$ \\
        
        \hfil A0514   &\hfil $5660\pm240$ &\hfil $5820\pm240$ &\hfil $73.5\pm0.6$&\hfil $0.12\pm0.001$&\hfil $0.59\pm0.01$, $0.26\pm0.01$&\hfil $0.105\pm0.002$ - $0.123\pm0.002$\\ 
         \hfil A1001   &\hfil $5100\pm190$  &\hfil $4960\pm180$&\hfil $72.4\pm0.7$&\hfil $0.138\pm0.003$&\hfil$0.56\pm0.01$, $0.24\pm0.01$&\hfil $0.129\pm0.002$ - $0.155\pm0.002$  \\ 
          \hfil V396 Lup &\hfil $5760\pm230$  &\hfil $5670\pm230$ &\hfil $78.0\pm0.4$&\hfil $0.133\pm0.002$&\hfil$0.58\pm0.01$, $0.26\pm0.01$&\hfil $0.110\pm0.002$ - $0.130\pm0.002$ \\ 
          \hfil A1707 &\hfil $6530\pm210$ &\hfil$6080\pm190$&\hfil $80.8\pm0.6$&\hfil $0.072\pm0.002$&\hfil $0.62\pm0.01$, $0.21\pm0.01$&\hfil $0.072\pm0.002$ - $0.082\pm0.002$ \\ 
           \hfil A1846  & \hfil $5380\pm190$ &\hfil$5720\pm210$&\hfil $87.0\pm0.7$&\hfil $0.150\pm0.002$&\hfil $0.56\pm0.01$, $0.25\pm0.01$&\hfil $0.117\pm0.001$ - $0.140\pm0.002$ \\
           \hfil A2022 &\hfil $6130\pm140$ & \hfil $6260\pm150$&\hfil $80.1\pm0.7$&\hfil$0.100\pm0.003$&\hfil $0.60\pm0.01$, $0.24\pm0.01$&\hfil $0.087\pm0.002$ - $0.100\pm0.002$ \\
        \hfil A2044  &\hfil $6010\pm80$  &\hfil $5940\pm80$&\hfil $87.2\pm0.6$&\hfil $0.088\pm0.001$&\hfil $0.61\pm0.01$, $0.22\pm0.01$&\hfil $0.084\pm0.003$ - $0.097\pm0.004$  \\ 
        \hfil A2132  &\hfil $6030\pm190$ &\hfil $6040\pm190$&\hfil $73.4\pm0.5$&\hfil $0.113\pm0.002$&\hfil $0.58\pm0.01$, $0.23\pm0.01$&\hfil $0.096\pm0.001$ - $0.112\pm0.002$  \\
        \hfil S2213   &\hfil $6130\pm120$  &\hfil $5960\pm120$&\hfil $90.00^{+0.00}_{-1.5}$&\hfil $0.088\pm0.003$&\hfil $0.61\pm0.01$, $0.22\pm0.01$&\hfil $0.081\pm0.003$ - $0.093\pm0.003$\\ 
          \hfil A2258  &\hfil $5990\pm50$ &\hfil $5970\pm50$&\hfil $90.00^{+0.00}_{-0.6}$&\hfil $0.087\pm0.002$&\hfil $0.57\pm0.01$, $0.22\pm0.01$&\hfil$0.081\pm0.003$ - $0.098\pm0.002$\\ 
          \hfil A2348 &\hfil $6190\pm230$ &\hfil $6070\pm220$&\hfil $73.9\pm0.5$&\hfil $0.117\pm0.002$&\hfil $0.59\pm0.01$, $0.24\pm0.01$&\hfil $0.087\pm0.003$ - $0.100\pm0.003$\\ \hline
           
    \end{tabular}
    \caption{Light curve solution summary of the main pertinent parameters. ($r_{1,2}$) = geometric fractional radii and ($q_{inst}$) = the instability mass ratio range.}
    \end{table*}

\section{Density and Chromospheric Activity}
\subsection{Component Density}

It is well known that the secondary components are larger than their main sequence counterparts. In addition to change in the radius some researchers \citep{2013MNRAS.430.2029Y} postulate that distortions of the secondary may result from it actually being initially a larger primary that has lost mass to the current primary and due to retention of higher mass core constituents is likely to be denser than the primary. \citet{2004A&A...414..317K} argues that the densities of the components will always be different and that the difference between density of the primary ($\rho_1$) and that of the secondary ($\rho_2$) will always be negative.

 As noted by \citep{1981ApJ...245..650M} the density of the components in ($\rm gcm^{-3}$) can be expressed as a function of the period, relative radii and mass ratio. The density difference ($\Delta\rho$) between the components can be expressed as:
 
 \begin{equation}
    \Delta\rho = \frac {0.0189}{r_1^3(1+q)P^2} - \frac {0.0189q}{r_2^3(1+q)P^2} 
\end{equation}

All our system confirm the finding that the secondary is much denser and that $\Delta\rho$ is negative. The findings (to two decimal places) are summerised in Table 6

\subsection {Chromospheric Activity}

Contact binary systems typically have synchronised rotation with short periods of less than 1 day. Such rapid rotation rates are thought to increase magnetic activity related to contact binary systems \citep{2019BlgAJ..31...97G}. Increased magnetic activity may manifest itself as magnetic stellar wind and magnetic breaking leading to loss of angular momentum from the system \citep{2004MNRAS.355.1383L}. Although it is difficult to directly measure angular momentum loss there are potential secondary indicators of increased magnetic activity, such as strong ultraviolet emissions and/or increased chromospheric activity \citep{2004MNRAS.355.1383L, 1983MNRAS.202.1221R, 1983HiA.....6..643V}. There are some chromospheric and magnetic signals associated with contact binaries with the most readily observed light curve feature being the asymmetry in the maxima (O’Connell effect). Apart from the O'Connell effect and incorporation of starspots, the analysis of light curves provides little indication of chromospheric activity. None of the systems described in this report show significant variation of maxima, however, that does not exclude the presence of significant magnetic and chromospheric activity. High energy spectral emissions provide a much clearer indicators of enhanced chromospheric/magnetic activity. In low mass dwarfs the chromospheric emissions are  by photospheric light in the visual band. Higher energy emissions particularly in the far-ultraviolet region provide a less obscured alternative \citep{2010PASP..122.1303S}. The GALEX (Galaxy Evolution Explorer) satellite imaged the sky in the far-ultraviolet band (FUV) centered on 1539 {\AA} and this can be employed to explore chromospheric activity of contact binaries\citep{2010PASP..122.1303S}. 

The $R_{\rm HK}^{\prime}$ index \citep{1984ApJ...279..763N} is an excepted measure of chromospheric emission strength with $\log R_{\rm HK}^{\prime} \geq -4.75$ characteristic of a more active star  \citep{1996AJ....111..439H}. \citet{2010PASP..122.1303S} matched GALEX FUV magnitudes ($m_ {\rm FUV}$) to the $\log R_{\rm HK}^{\prime}$ for dwarf stars to derive the $\Delta(m_{\rm FUV-B})$ colour excess:
\begin{equation}
    \Delta(m_{\rm FUV-B}) = (m_{\rm FUV}-B) - (m_{\rm FUV}-B)_{\rm base}
\end{equation}
where
\begin{equation}
    (m_{ \rm FUV}-B)_{\rm base} = 6.73(B-V) + 7.43,
\end{equation}
They deduced that for active stars the colour excess was always below -0.5 and often less than -1.0. Less active stars usually had colour excess well above -0.5.

Six of our twelve systems were observed by the GALEX mission with recorded FUV magnitudes. We calculated the colour excess for all six using the above relationships and show that all have an ultraviolet colour excess well below -0.5 indicative of significant chromospheric/magnetic activity in the absence of light curve features. Magnetically active features, sometimes referred to as plages \citep{2008LRSP....5....2H}, are responsible for the chromospheric emissions while starspots are regions of intense magnetic activity leading to suppression of convection in the photosphere \citep{2003A&ARv..11..153S}. Although starspots are accompanied by plague regions, the reverse is not always true such that chromospheric activity is possible without photospheric starspots \citep{2017ApJ...835..158M}. The UV colour excesses are summarised in Table 6.

Coronal and chromospheric activity combined with fast synchronous rotation of the common envelopes may also result in X-Ray emissions from contact binaries \citep{2015MNRAS.446..510K, 2004A&A...415.1113G}. Seven of our systems (A0458, A0514, V396 Lup, A1846, A2044, S2213, A2348) were reported as potential X-Ray sources \citep{ 2022A&A...663A.115L} including three (A0514, V396 Lup and A2044) without FUV observations bringing to a total of nine systems with signs suggestive of increased chromospheric activity without characteristic photospheric changes. 

\begin{table}

   \centering

   \begin{tabular}{|l|l|l|l|l|}
    \hline
        \hfil Name &\hfil $\rho_1(\rm gcm^{-3})$ &\hfil $\rho_2(\rm gcm^{-3})$&\hfil $\Delta\rho$ &\hfil$\Delta (m_{\rm {FUV}}-B)$  \\
        \hline
        \hfil A0458  &\hfil $0.72\pm0.01$  &\hfil $1.49\pm0.01$&\hfil $-0.77$&\hfil $-3.34$\\
        
        \hfil A0514    &\hfil $0.69\pm0.01$ &\hfil $0.96\pm0.01$&\hfil $-0.27$&\hfil $--$ \\ 
         \hfil A1001    &\hfil $1.21\pm0.02$&\hfil $2.13\pm0.02$&\hfil $-0.92$&\hfil$--$  \\ 
          \hfil V396 Lup   &\hfil $0.65\pm0.03$ &\hfil $0.96\pm0.02$&\hfil $-0.31$&\hfil$--$ \\ 
          \hfil A1707 & \hfil$0.27\pm0.01$&\hfil $0.50\pm0.02$&\hfil $-0.23$&\hfil $--$ \\ 
           \hfil A1846  &  \hfil$1.02\pm0.01$&\hfil $1.72\pm0.02$&\hfil $-0.70$&\hfil $--$ \\
           \hfil A2022  & \hfil $0.67\pm0.02$&\hfil $1.04\pm0.02$&\hfil$-0.37$&\hfil $-2.30$ \\
        \hfil A2044   &\hfil $0.56\pm0.01$&\hfil $1.05\pm0.01$&\hfil $-0.49$&\hfil $--$  \\ 
        \hfil A2132   &\hfil $0.87\pm0.02$&\hfil $1.58\pm0.02$&\hfil $-0.71$&\hfil $-2.80$  \\
        \hfil S2213    &\hfil $0.56\pm0.01$&\hfil $1.05\pm0.02$&\hfil $-0.49$&\hfil $-2.32$ \\ 
          \hfil A2258  &\hfil $0.87\pm0.01$&\hfil $1.32\pm0.02$&\hfil $-0.45$&\hfil $-2.56$\\ 
          \hfil A2348  &\hfil $0.68\pm0.01$&\hfil $1.19\pm0.02$&\hfil $-0.51$&\hfil $-3.23$ \\ 
          \hline
           
    \end{tabular}
    \caption{Density and ultraviolet colour excess. }
    \end{table}

\section{Discussion and Conclusion}

Photometric observations and light curve solution of twelve low mass contact binary systems is presented. All were found to be of extreme low mass ratio ranging from 0.072 to 0.15. Compared to theoretical parameters all have a mass ratios indicative of likely orbital instability and are likely merger (red nova) candidates. Regular high cadence future observations are encouraged to accurately determine period variations and brightness changes which maybe an indicator of impending merger \citep{2011A&A...528A.114T}. A thorough study of orbital behavior of contact binaries have been performed recently by \citet{2020MNRAS.497.3493Z, 2022MNRAS.510.5315P, 2022MNRAS.514.5528L}. The first and the last  paper discuss in particular the systems close to the orbital period cut-of. Ultra-short period systems seem to have generally stable orbits and do not show evidence of significant period decrease. From the viewpoint of Darwin instability, this is understandable, since these are not such a low-mass ratio systems and it may take a long evolutionary time for ultra-short contact binaries to reach an extremely low mass ratio and become dynamically unstable \citep{2022MNRAS.514.5528L}.

Review of some astrophysical characteristics confirm similarity with other contact binary systems with the secondary considerably brighter and larger than main sequence stars of similar mass. In addition, the secondary in all cases is significantly denser than the current primary. Chromospheric activity in contact binary is usually detected as a variation in the two maxima from the light curve due to starspots. As we have shown in this study non-photospheric markers probably can detect such activity without typical light curve features. Another high energy feature of excess chromospheric/magnetic activity is X-Ray emissions \citep{2004A&A...415.1113G}. Only a small fraction of contact binary star systems are X-ray sources \citep{2022A&A...663A.115L} such that presence of X-Ray emissions can be taken as a marker of increased magnetic activity and potential indicator for an increased loss of angular momentum and orbital instability. Seven of twelve systems reported have been catalogued as having X-Ray emissions \citep{2022A&A...663A.115L}. Also, as noted above the FUV magnitude can be used as a marker for chromospheric activity. Another interesting point of note is how much brighter relative to main sequence stars are the contact binaries in the FUV band. \citet{2010AJ....139.1338F} have estimated the  absolute FUV magnitudes ($M_{\rm FMS}$) for single main sequence stars from B8 to M2 spectral types. For the six systems from our sample with FUV magnitude we find that all systems are between 0.5 and over 3.0 magnitude brighter in the FUV band relative to their main sequence counterparts.

Progress in the detection and monitoring of potential contact binary merger candidates is rapidly progressing. There now exists a theoretical framework to determine the potential for merger from the estimate of the mass of the primary and the mass ratio of the system \citep{2021MNRAS.501..229W}. Models and techniques have been developed to rapidly select potential merger candidates from survey photometry \citep{2022JApA...43...94W}. However, follow up dedicated observations are required due to the low resolution of most survey data \citep{2022arXiv221208209W}.

\begin{acknowledgements}

\noindent Acknowledgements.

Based on data acquired on the Western Sydney University, Penrith Observatory Telescope. We acknowledge the traditional custodians of the land on which the Observatory stands, the Dharug people, and pay our respects to elders past and present.\\

This research has made use of the SIMBAD database, operated at CDS, Strasbourg, France.\\

This research has made use of the VizieR catalogue access tool, CDS, Strasbourg, France (DOI : 10.26093/cds/vizier).\\

B. Arbutina, G. Djurašević and J. Petrović  acknowledge the funding provided by the Ministry of Science, Technological Development and Innovation of the Republic of Serbia through the contracts  451-03-47/2023-01/200104 (BA) and 451-03-47/2023-01/200002 (GDj, JP).


\end{acknowledgements}


\bibliography{sample631}{}
\bibliographystyle{aasjournal}



\end{document}